\documentclass[preprint,amsmath,amssymb,aps,showkeys,showpacs]{revtex4}
\usepackage[english]{babel}
\usepackage{graphicx}
\usepackage{graphics}
\usepackage{amsmath}
\usepackage{dcolumn}
\usepackage{amssymb}
\usepackage{bm}

\begin{document}
\title{Linear confinement of a scalar particle in a G\"odel-type spacetime}
\author{R. L. L. Vit\'oria}
\affiliation{Departamento de F\'isica, Universidade Federal da Para\'iba, Caixa Postal 5008, 58051-900, Jo\~ao Pessoa-PB, Brazil.}

\author{C. Furtado}
\email{furtado@fisica.ufpb.br}
\affiliation{Departamento de F\'isica, Universidade Federal da Para\'iba, Caixa Postal 5008, 58051-900, Jo\~ao Pessoa-PB, Brazil.}

\author{K. Bakke}
\email{kbakke@fisica.ufpb.br}
\affiliation{Departamento de F\'isica, Universidade Federal da Para\'iba, Caixa Postal 5008, 58051-900, Jo\~ao Pessoa-PB, Brazil.}

\begin{abstract}

Based on the studies of confinement of quarks, we introduce a linear scalar potential into the relativistic quantum dynamics of a scalar particle. Then, we analyse the linear confinement of a relativistic scalar particle in a G\"odel-type spacetime in the presence of a topological defect. We consider a G\"odel-type spacetime associated with null curvature, i.e., the Som-Raychaudhuri spacetime, which is characterized by the presence of vorticity in the spacetime. Then, we search for analytical solutions to the Klein-Gordon equation and analyse the influence of the topology of the cosmic string and the vorticity on the relativistic energy levels.

\end{abstract}

%\pacs{03.65.Pm, 03.65.Ge}

\maketitle

\section{Introduction}

The first solution to Einstein's equations that considers the rotation of a homogeneous mass distribution with cylindrical symmetry was given by G\"odel \cite{godel}.  This important and historical solution has a great interest since it was the first and the most well-known solution that describes a model of universe where the causality is violated \cite{sthephani}. This solution is characterized by the possibility of existing closed timelike curves (CTCs). Hawking \cite{hawking} studied the physical properties of the presence of CTCs in this geometry, and conjectured that the presence of CTCs is physically inconsistent. Rebou\c cas {\it et al} \cite{reboucas,reboucas2,reboucas3} investigated the G\"odel-type solutions and the possible sources of the model in the context of general relativity. They showed that the G\"odel solution can be generalized to cylindrical coordinates. Besides, Rebou\c cas {\it et al} \cite{reboucas,reboucas2,reboucas3} analysed the problem of causality, and then, established three classes of solutions that are characterized by the following properties: (i) solutions with no CTCs; (ii) solutions with a sequence of causal regions and not causal regions that alternates with each other; (iii) solutions characterized by only one non-causal region. 

In recent years, investigations of relativistic quantum effects on scalar and spin-half particles in G\"odel-type spacetimes have been addressed by several authors. In Ref. \cite{figueiredo}, the first study of this problem is made, where it is investigated the Klein-Gordon and Dirac equations in the G\"odel-type spacetimes with positive and negative curvatures, and also in the flat G\"odel-type spacetime. Further, the investigation of the relationship between the quantum dynamics of the scalar quantum particle in background of general relativity with G\"odel solutions and the Landau levels in the flat, spherical and hyperbolic spaces was performed by Drukker {\it et al} \cite{fiol}. They analyzed the similarity between the spectrum of energy of scalar quantum particle in these class of G\"odel-type spacetimes and the Landau levels in curved backgrounds. This similarity has also been observed by Das and Gegenberg \cite{gengeberge} by studying the Klein-Gordon equation in the Som-Raychaudhuri spacetime (G\"odel flat solution). They also compared with the Landau levels in the flat space. Recently, the scalar quantum particle has been investigated in the class of the G\"odel solutions with a cosmic string passing through the spacetime \cite{j2}.

In this work, we consider a G\"odel-type spacetime in the presence of a topological defect, and then, analyse the linear confinement of a relativistic scalar particle. In recent years, several authors have investigated the physical properties of a series of backgrounds with a cosmic string. For example, the Schwarzschild spacetime with a cosmic string \cite{mukunda,germano}, the Kerr spacetime with a cosmic string \cite{22}, G\"odel-type spacetime with a cosmic string \cite{j2}  and the cosmic string in AdS space \cite{demellosahar}. Furthermore, it is worth mentioning the studies of a scalar quantum particle confined in two concentric thin shells in curved spacetime backgrounds with a cosmic string \cite{sandro}, the Klein-Gordon oscillator in a Som-Raychaudhuri spacetime with a cosmic dispiration \cite{j1}, and fermions in a family of G\"odel-type solutions with a cosmic string \cite{gabriel}. It is worth observing that the investigation of the influence of the topology of the cosmic string spacetime on the linear confinement of the scalar particle made in this work can be useful in studying the effects of disclinations (an analogue of the cosmic string) on condensed matter systems. An example is the quantum Hall effect in the presence of rotation and disclination, where  a linear interaction is considered. Therefore, in this work, we shall deal with a G\"odel-type spacetime with null curvature, i.e., in the Som-Raychaudhuri spacetime \cite{som}. We also consider the presence of the cosmic string. Thereby, we search for relativistic bound states solutions, where we discuss the influence of the spacetime background on the relativistic energy levels.

This paper is organized as follows: in section II, we introduce the Som-Raychaudhuri spacetime with the cosmic string. Then, we obtain the Klein-Gordon equation that describes the confinement of a scalar particle to a linear scalar potential. We show that we can solve the Klein-Gordon equation analytically, and thus we discuss the influence of the spacetime background on the relativistic energy levels of the system; in section III, we present our conclusions.

\section{Linear confinement in the Som-Raychaudhuri spacetime with a cosmic string}

In this work, we are interested in a particular case of the G\"odel-type solutions. We wish to analyse the linear confinement of a scalar particle in a G\"odel-type spacetime by searching for relativistic bound states. This particular solution is called as the Som-Raychaudhuri spacetime with a the cosmic string \cite{j1,j2}. In recent decades, this type of solution has been used in studies of string theory \cite{horo,russo,russo1}. Thus, the Som-Raychaudhuri spacetime with a cosmic string is described by the line element (with $c=\hbar=1$):
\begin{eqnarray}
ds^2=-\left(dt+\alpha\,\Omega\,r^{2}d\varphi\right)^{2}+\alpha^{2}r^{2}d\varphi^{2}+dr^{2}+dz^{2}.
\label{1.1}
\end{eqnarray}
The parameter $\Omega$ characterizes the vorticity of the spacetime, while the parameter $\alpha$ characterizes the cosmic string. In short, the parameter $\alpha$ is associated with the deficit of angle $\alpha=(1-4\lambda)$, where $\lambda$ is the mass per unit length of the cosmic string. It has values in the range $0<\alpha<1$. Observe that the line element defined above can be written as
\begin{eqnarray}
ds^2=-(dt + A_{i}dx^{i})^{2} + h_{ij}dx^{i}dx^{j},
\end{eqnarray}
where the spatial coordinates of the spacetime are represented by $x^{i}$. Note that in case of the Som-Raychaudhuri spacetime, we have the form
\begin{eqnarray}
ds^2=-(dt + \Omega(ydx-xdy))^{2} + dx^{2} + dy^{2} + dz^{2},
\end{eqnarray}
where we have used the cartesian coordinates $(t,x,y,z)$ because it is clearer to observe the similarity. From Eq. (\ref{1.1}), we obtain $A_{\varphi}=\alpha\,\Omega\,r^{2}$ in cylindrical coordinates. This expression is analogous to the vector potential that yields a uniform magnetic field in $z$-direction. An interesting property of this type of metric is that the geodesics for this spacetime are circles, which have a physical description similar to Larmor orbits for electron that moves in a perpendicular magnetic field \cite{fiol,gengeberge}. This analogy also arises from the point of view of quantum mechanics, where the quantum dynamics of scalar and spinorial quantum particles in this spacetime is an analogue of the Landau levels as observed in Refs. \cite{fiol,gengeberge,j1,j2,gabriel}. We shall discuss this point later.

On the other hand, the linear confinement of quantum particles has a great importance for models of confinement of quarks \cite{linear1}. It is worth pointing out that the linear scalar potential has attracted a great interest in atomic and molecular physics \cite{linear3a,linear3b,linear3c,linear3d,linear3e,linear3f,fb} and in relativistic quantum mechanics \cite{linear2,linear2a,linear2b,linear2c,linear2d,linear2e,linear2f,vb,vb2,vb3}. Our objective is to investigate the effects of the background defined in Eq. (\ref{1.1}) on the linear confinement of a relativistic scalar particle. By following Ref. \cite{greiner}, we introduce the scalar potential into relativistic wave equation through the mass term: $M\rightarrow M+V\left(\vec{r},\,t\right)$, where $M$ is the mass of the free particle and $V\left(\vec{r},\,t\right)$ is the scalar potential. Hence, let us consider the linear scalar potential \cite{linear1}: $V\left(r\right)=\kappa\,r$, where $\kappa$ is a constant. Thereby, the Klein-Gordon equation that describes the linear confinement of a scalar particle in the Som-Raychaudhuri spacetime with the cosmic string is given by
\begin{eqnarray}
-\frac{\partial^{2}\phi}{\partial t^2}+\frac{1}{r}\frac{\partial}{\partial r}\left(r\frac{\partial\phi}{\partial r}\right)+\left(\frac{1}{\alpha r}\frac{\partial}{\partial \varphi}-\Omega r\frac{\partial}{\partial t}\right)^{2}\phi+\frac{\partial^{2}\phi}{\partial z^{2}}-\left(M+\kappa\,r\right)^2\phi=0.
\label{1.3}
\end{eqnarray}

Observe that this system has the cylindrical symmetry, therefore, the eigenvalues of the $z$-component of the angular momentum operator and the $z$-component of the linear momentum operator are conserved quantities. In this way, we can write the solution to Eq. (\ref{1.3}) as $\phi\left(t,\,r,\,\varphi,\,z\right)=e^{-i\mathcal{E}\,t}\,e^{i\,l\,\varphi}\,e^{ip_{z}\,z}\,u\left(r\right)$, where $l=0,\pm1,\pm2,\pm3,\ldots$ are the eigenvalues of the $z$-component of the angular momentum operator and $p_{z}=\mathrm{const}$ are the eigenvalues of the $z$-component of the linear momentum operator. In this way, we obtain the following equation for $u\left(r\right)$:
\begin{eqnarray}
u''+\frac{1}{r}\,u'-\frac{l^{2}}{\alpha^{2}\,r^{2}}\,u-\omega^{2}\,r^{2}\,u-2M\kappa\,r\,u+\left[\mathcal{E}^{2}-M^{2}-p_{z}^{2}-\frac{2\Omega\,l\,\mathcal{E}}{\alpha}\right]u=0,
\label{1.4}
\end{eqnarray}
where $\omega=\sqrt{\Omega^{2}\mathcal{E}^{2}+\kappa^{2}}$. Let us define $x=\sqrt{\omega}\,r$, then, Eq. (\ref{1.4}) becomes
\begin{eqnarray}
u''+\frac{1}{x}\,u'-\frac{l^{2}}{\alpha^{2}\,x^{2}}\,u-x^{2}\,u-\theta\,x\,u+\beta\,u=0,
\label{1.5}
\end{eqnarray}
where we have defined the parameters $\omega$ and $\beta$ in the above equation as 
\begin{eqnarray}
\theta=\frac{2M\kappa}{\omega^{3/2}};\,\,\,\,\beta=\frac{1}{\omega}\left[\mathcal{E}^{2}-M^{2}-p_{z}^{2}-\frac{2\Omega\,l\,\mathcal{E}}{\alpha}\right].
\label{1.6}
\end{eqnarray}

Next, we use the appropriated boundary conditions to investigate the bound states in this problem. It is required that the regularity of wave function at the origin and the normalizability at infinity. Then, we proceed with the analysis of the asymptotic behaviour of the radial eigenfunction at origin and in the infinite. These conditions are necessary since the wave function must be well-behaved in this limits, and thus, bound states of energy for the linear confinement can be obtained. Note that these limits correspond to the critical points of the Eq. (\ref{1.5}). With these conditions, we obtain the convergence of the wave function at origin ($x\rightarrow0$) and at infinite ($x\rightarrow\infty$). Let us impose that $u\left(x\right)\rightarrow0$ when $x\rightarrow0$ and $x\rightarrow\infty$, hence, the solution to Eq. (\ref{1.6}) is given by
\begin{eqnarray}
u\left(x\right)=x^{\left|l\right|/\alpha}\,e^{-\frac{x^{2}}{2}}\,e^{-\frac{\theta}{2}\,x}\,H\left(x\right),
\label{1.7}
\end{eqnarray}
where $H\left(x\right)$ is the solution to the biconfluent Heun equation \cite{heun}, which has the form:
\begin{eqnarray}
H''+\left[\frac{\frac{2\left|l\right|}{\alpha}+1}{x}-\theta-2x\right]H'+\left[\beta+\frac{\theta^{2}}{4}-2-2\left|l\right|-\frac{\theta\left(\frac{2\left|l\right|}{\alpha}+1\right)}{2x}\right]H=0.
\label{1.8}
\end{eqnarray}

With the goal of obtaining analytical solutions to Eq. (\ref{1.8}), let us write $H\left(x\right)=\sum_{j=0}^{\infty}\,d_{j}\,x^{j}$, then, from Eq. (\ref{1.8}) we obtain the relation:
\begin{eqnarray}
d_{1}=\frac{\theta}{2}\,d_{0},
\label{1.9}
\end{eqnarray}
and the recurrence relation:
\begin{eqnarray}
d_{j+2}=\frac{\theta\left(2j+3+\frac{2\left|l\right|}{\alpha}\right)}{2\left(j+2\right)\left(j+2+\frac{2\left|l\right|}{\alpha}\right)}\,d_{j+1}-\frac{\left(4\beta+\theta^{2}-8-\frac{8\left|l\right|}{\alpha}-8j\right)}{4\left(j+2\right)\left(j+2+\frac{2\left|l\right|}{\alpha}\right)}\,d_{j}.
\label{1.10}
\end{eqnarray}

A polynomial solution to $H\left(x\right)$ is achieved when we impose that the series terminates. Naturally, there is a convergence problem in this solution that can be solved when we impose: 
\begin{eqnarray}
4\beta+\theta^{2}-8-\frac{8\left|l\right|}{\alpha}=8n;\,\,\,\,\,d_{n+1}=0.\,\,\,\,\left(n=1,2,3,\ldots\right)
\label{1.11}
\end{eqnarray}
Hence, there are two conditions that must be satisfied in order that the series terminates. Note that $n=1,2,3,\ldots$ is the quantum number associated with the radial modes. From the condition $4\beta+\theta^{2}-8-\frac{8\left|l\right|}{\alpha}=8n$, we obtain 
\begin{eqnarray}
\mathcal{E}_{n,l}^{2}-\frac{2\Omega\,l}{\alpha}\,\mathcal{E}_{n,l}-C_{n,\,l}=0,
\label{1.12}
\end{eqnarray}
where
\begin{eqnarray}
C_{n,\,l}=M^{2}+p_{z}^{2}+2\omega\left(n+\frac{\left|l\right|}{\alpha}+1\right)-\frac{\kappa^{2}M^{2}}{\omega^{2}}.
\label{1.12a}
\end{eqnarray}
%As we shall see later, Eq. (\ref{1.12}) can be used to determine the energy levels of the system. 

However, if we wish to obtain a polynomial solution to $H\left(x\right)$, for example, a polynomial of first degree ($n=1$), therefore, from the condition $d_{n+1}=0$, we have that $d_{2}=0$, and thus, we obtain the relation: 
\begin{eqnarray}
\omega_{1,\,l}=\left(\frac{M^{2}\kappa^{2}_{1,\,l}}{2}\left[3+2\frac{\left|l\right|}{\alpha}\right]\right)^{1/3}.
\label{1.13}
\end{eqnarray} 
The relation (\ref{1.13}) is obtained by assuming that the parameter $\kappa$ can be adjusted, and thus, the polynomial of first degree can be achieved \cite{eug,vb3}. In this way, by substituting Eq. (\ref{1.13}) into Eq. (\ref{1.12}), we obtain the second degree algebraic equation for $\mathcal{E}_{1,\,l}$:
\begin{eqnarray}
\mathcal{E}_{1,\,l}^{2}-\frac{2\Omega\,l}{\alpha}\mathcal{E}_{1,\,l}-C_{1,\,l}=0,
\label{1.14}
\end{eqnarray}
where 
\begin{eqnarray}
C_{1,\,l}=M^{2}+p_{z}^{2}+2\left[2+\frac{\left|l\right|}{\alpha}\right]\left(\frac{M^{2}\kappa^{2}_{1,\,l}}{2}\left[3+2\frac{\left|l\right|}{\alpha}\right]\right)^{1/3}-\frac{M^{2}\kappa^{2}_{1,\,l}}{\left(\frac{M^{2}\,\kappa^{2}_{1,\,l}}{2}\left[3+2\frac{\left|l\right|}{\alpha}\right]\right)^{2/3}}.
\label{1.15}
\end{eqnarray}

Hence, the solutions to Eq. (\ref{1.14}) yield the allowed energies associated with the radial mode $n=1$:
\begin{eqnarray}
\mathcal{E}_{1,\,l}=\frac{\Omega\,l}{\alpha}\pm\sqrt{C_{1,\,l}+\frac{\Omega^{2}\,l^{2}}{\alpha^{2}}}.
\label{1.16}
\end{eqnarray}

By comparing with the expression for the relativistic energy level of the ground state $\left(n=1;\,l=0\right)$ in the Minkowski spacetime discussed in Ref. \cite{vb3}, we have that the vorticity of the spacetime and deficit of angle modify the relativistic energy level of the ground state. In the Som-Raychaudhuri spacetime with a cosmic string, the ground state of the linear confinement of a relativistic scalar particle has two allowed energies as shown in Eq. (\ref{1.16}). Observe that the topology of the cosmic string modifies the angular momentum operator and gives an effective angular momentum quantum number $l_{\mathrm{eff}}=l/\alpha$, i.e., a fractional angular momentum. Furthermore, with both conditions established in Eq. (\ref{1.11}) satisfied, for $n=1$, the function $H\left(x\right)$ is given by a polynomial of first degree, therefore, the wave function (\ref{1.7}) is written in the form:  
\begin{eqnarray}
u_{1,\,l}\left(x\right)=x^{\left|l\right|/\alpha}\,e^{-\frac{x^{2}}{2}}\,e^{-\frac{\theta}{2}\,x}\,\left(1+\frac{\theta}{2}\,x\right),
\label{1.16a}
\end{eqnarray}
which is associated with the ground state of the system. 

Besides, a particular case is given by taking $\alpha=1$ in Eq. (\ref{1.16}), which yields 
\begin{eqnarray}
\bar{\mathcal{E}}_{1,\,l}=\Omega\,l\pm\sqrt{\bar{C}_{1,\,l}+\Omega^{2}\,l^{2}},
\label{1.17}
\end{eqnarray}
where 
\begin{eqnarray}
\bar{C}_{1,\,l}=M^{2}+p_{z}^{2}+2\left[2+\left|l\right|\right]\left(\frac{M^{2}\kappa^{2}_{1,\,l}}{2}\left[3+2\left|l\right|\right]\right)^{1/3}-\frac{M^{2}\,\kappa^{2}_{1,\,l}}{\left(\frac{M^{2}\,\kappa^{2}_{1,\,l}}{2}\left[3+2\left|l\right|\right]\right)^{2/3}}.
\label{1.18}
\end{eqnarray}
Therefore, Eq. (\ref{1.17}) corresponds to the allowed energies of the linear confinement of the relativistic scalar particle in the Som-Raychaudhuri spacetime. In this case, only the vorticity of the spacetime modifies the ground state energy in contrast to that discussed in Ref. \cite{vb3}. 

Note that, in the limit $\kappa=0$ (i.e., in the absence of the linear scalar potential), we obtain from Eq. (\ref{1.3}) to Eq. (\ref{1.8}) the results obtained in Ref. \cite{j1} for a scalar particle in the Som-Raychaudhuri spacetime with a cosmic string. In Refs. \cite{j1,fiol, gengeberge}, it is demonstrated the closed similarity between the energy levels of a scalar quantum particle in the Som-Raychaudhuri spacetime and the Landau levels \cite{landau}. Observe that, in this limit, Eq. (\ref{1.8}) becomes the hypergeometric equation. This limit corresponds to taking $\theta=0$ in Eq. (\ref{1.11}), then, we obtain only the first relation of Eq. (\ref{1.11}): 
\begin{eqnarray}
4\beta-8-\frac{8\left|l\right|}{\alpha}=8n'; \,\,\,\,\left(n'=0,1,2,3,\ldots\right),
\label{1.11a}
\end{eqnarray}
and thus, one of the eigenvalues of the energy is
\begin{eqnarray}
\mathcal{E}=\left(n' +\frac{\left|l\right|}{\alpha} +\frac{l}{\alpha}+1\right)\Omega +\sqrt{\left(n' +\frac{\left|l\right|}{\alpha} +\frac{l}{\alpha}+1\right)^{2}\Omega^{2} + p_{z}^{2} + M^{2}}.
\label{1.11c}
\end{eqnarray}

Note that the eigenvalue (\ref{1.11c}) is the same that obtained for Som-Raychaudhuri spacetime with a cosmic string in Ref. \cite{j1}, and also by taking the limit $\alpha=1$ in Refs. \cite{fiol,gengeberge}. The eigenvalue (\ref{1.11c}) is similar to the Landau levels for scalar particle as demonstrated in Refs. \cite{j1,fiol, gengeberge}, where the rotation plays the role of the uniform magnetic field and the parameter $\Omega$ plays the role of the cyclotron frequency. This similarity is more visible in the limit case where $M=p_{z}=0$, where the eigenvalue (\ref{1.11c}) becomes $\mathcal{E}=2\Omega\left(n'+\frac{\left|l\right|}{\alpha} +\frac{l}{\alpha}+1\right)$. This is similar to the eigenvalue obtained for the Landau levels in the presence of a cosmic string in Ref. \cite{bfmv}. This similarity is the central point to claim the use of the Som-Raychaudhuri spacetime to investigate the linear confinement of quarks in the presence of magnetic field and topological defect.

Returning to Eq. (\ref{1.16}), it is interesting to observe that bound states exist only for restrict values of the linear confinement constant $\kappa$, as we can observe from Eq. (\ref{1.13}) to Eq. (\ref{1.16}). This fact results from the series $H\left(x\right)=\sum_{j=0}^{\infty}\,d_{j}\,x^{j}$ to be required to terminate. Note that the problem which is investigated in this work is very different to that investigated in Refs. \cite{vb,vb2,vb3}. Here, we investigate the relativistic spinless quantum particle in a rotating spacetime in the presence of a topological defect. We recover the results of Refs. \cite{vb,vb2,vb3} in the appropriated limits. In this way, the results (\ref{1.16}) correspond to the bound state associated with the radial mode $n=1$, where the allowed energies depend on the rotation parameter $\Omega$ and the parameter $\alpha$ that characterizes the cosmic string. It is interesting to observe that the behaviour of the Klein-Gordon particle in presence of the linear interaction in the Som-Raychaudhuri spacetime is analogous to the Landau levels in the presence of a linear scalar potential. In the present case, the rotation plays the role of external magnetic field. The influence of the parameter $\Omega$ can be viewed as the influence of the external field on the bound states, therefore, these physical results can be used to investigate linear confinement of quarks.

Despite our discussion has focused on the radial mode $n=1$, by following the steps from Eq. (\ref{1.11}) to Eq. (\ref{1.14}), hence, we can obtain the allowed energies associated with the radial modes $n=2$, $n=3$ and so on.

\section{Conclusions}

We have investigated the influence of the vorticity and a topological defect on the linear confinement of a relativistic scalar particle. The vorticity and the topological defect stem from a particular G\"odel-type solution called as the Som-Raychaudhuri spacetime with a cosmic string \cite{j1,j2}. By analysing the energy associated with the radial mode $n=1$, therefore, we have seen that both vorticity and the topology of the cosmic string modify this energy level and give rise to the allowed energies written in Eq. (\ref{1.16}). Moreover, the topology of the cosmic string gives rise to an effective angular momentum $l_{\mathrm{eff}}=l/\alpha$. A particular case of the allowed energies associated with the radial mode $n=1$ is given by taking $\alpha=1$, then, we have obtained the expression of the allowed energies in the Som-Raychaudhuri spacetime as shown in Eq. (\ref{1.17}). In this case, we have observed that only the vorticity modifies the ground state energy. 

In this paper, we have obtained the bound states for the linear confinement of a scalar quantum particle in the Som-Raychaudhuri spacetime with a topological defect and analysed the similarities between the present results and the with Landau levels in this spacetime background. As shown in Refs. \cite{j1,fiol, gengeberge}, due to the closed relation of the Landau levels to the relativistic quantum energy levels in this G\"odel-type universe, this result can be used to discuss the Landau levels in condensed matter systems of interacting particles. This closed relation can also be used in solving problems of the linear confinement of quarks in the presence of a topological defect, where a magnetic field can be introduced in a geometric way through the Som-Raychaudhuri spacetime metric. We claim that the results obtained here can be used to investigate heavy quarkonia subject to a strong magnetic field \cite{bonati} in the presence of topological defect, due to the similarity between the quantum dynamics in Som-Raychaudhuri spacetime and the Landau Levels \cite{j1,fiol, gengeberge}. Models with confining potentials have been used to describe the spectrum of quarkonia-type systems. In the case of the linear confining potential, this would be the confining potential of quarks. This type of potential has been used to describe quarkonia with heavy quark. Thereby, the study of the linear confinement in the Som-Raychaudhuri spacetime in the presence of a topological defect can be used as a model, where the rotation plays the role of an external magnetic field. Therefore, the interaction is introduced by the potential $V=\kappa\,r$ and the geometric background, in turn, introduces a ``magnetic field'' (it arises from the vorticity of the spacetime). As a future study, we can mention the introduction of the Coulomb interaction plus the linear potential, hence, we could investigate the Cornell potential \cite{cornell} in the flat G\"odel-type universe.

\acknowledgments{We would like to thank the Brazilian agencies CNPq and CAPES for financial support.}

\end{document}